# On the Merging of Domain-Specific Heterogeneous Ontologies using Wordnet and Web Pattern-based Queries

Mohammed Maree
Faculty of Information Technology
Monash University
mohammed.maree@infotech.monash.edu.my

Mohammed Belkhatir
Faculty of Computer Science
University of Lyon
mohammed.belkhatir@univ-lyon1.fr

**ABSTRACT**

Ontologies form the basic interest in various computer science disciplines such as semantic web, information retrieval, database design, etc. They aim at providing a formal, explicit and shared conceptualization and understanding of common domains between different communities. In addition, they allow for concepts and their constraints of a specific domain to be explicitly defined. However, the distributed nature of ontology development and the differences in viewpoints of the ontology engineers have resulted in the so called "semantic heterogeneity" between ontologies. Semantic heterogeneity constitutes the major obstacle against achieving interoperability between ontologies. To overcome this obstacle, we present a multi-purpose framework which exploits the WordNet generic knowledge base for: *i)* Discovering and correcting the incorrect semantic relations between the concepts of the ontology in a specific domain. This step is a primary step of ontology merging. *ii)* Merging domain-specific ontologies through computing semantic relations between their concepts. *iii)* Handling the issue of *missing concepts* in WordNet through the acquisition of statistical information on the Web. *And iv)* Enriching WordNet with these missing concepts. An experimental instantiation of the framework and comparisons with state-of-the-art syntactic and semantic-based systems validate our proposal.

## 1. INTRODUCTION

Ontologies are means of conceptually structuring domain knowledge. They are being built in order to formally and explicitly represent our conceptualizations in common domains of interest. However, due to the distributed nature of ontology development and different viewpoints of ontology engineers even about the same domain, semantic heterogeneity between ontologies becomes a critical issue. Ontologies can be heterogeneous at two different levels of analysis. The first case of heterogeneity is related to the languages used to create ontologies. The reason is that, ontologies are created by different languages and represented using different syntaxes. Language heterogeneity is not our research issue since we have tools[1] for converting ontology languages from one into another. The second case of heterogeneity is at the ontology component level and known as the "semantic heterogeneity". In this case, conflicts in the semantic relations and logical inconsistencies or semantic mismatches may exist between the concepts of the ontologies. For example, we may have two or more domain-specific ontologies which use different terms to describe the same concept or use the same term to describe different concepts.

To solve the semantic heterogeneity problem and achieve interoperability between ontologies, we need to resolve the conflicts and logical inconsistencies between them. This process can be done through merging two or more source ontologies from the same domain into a single coherent ontology, or through finding alignments among them by establishing links between their entities and allowing information to be reused among them. Merging or aligning methods exploit several matching strategies such as lexical-based techniques like the Levenshtein distance technique [9] which is used in [12], structure-based approaches [4] [5], strategies that use an external resource [7] or a combination of these [3] [10] in order to find semantic correspondences between the concepts of the ontologies.

---

[1] http://owl.cs.manchester.ac.uk/convertor/

Manual ontology merging is a difficult, time-consuming and error prone task due to the continuous growth in both the size and number of ontologies. Therefore, several automatic and semi-automatic ontology merging frameworks have recently been proposed. These frameworks aim at finding semantic correspondences between the concepts of the ontologies for a specific domain through exploiting syntactic-based, semantic-based, or both techniques.

Existing approaches suffer from a number of problems. First, they rely on linguistic-based and structure-based similarity measures. Although these techniques may produce good results, there is a considerable number of cases where they fail to discover mappings. For example, the relation between the concepts (student and organism) may not be discovered using such techniques. To overcome this drawback, some of the approaches propose the integration of an external knowledge base to support the process of finding semantically corresponding concepts of the source ontologies. However, problems may exist at the level of the used knowledge base. For instance, we may find background information such as concepts names or instances missing in the knowledge base. As for those approaches which use WordNet [14] as a knowledge base, we may find concepts such as "Corporate Body" or instances such as "Stanford University" missing. Therefore, the merging or alignment process will be negatively affected.

We propose an ontology merging framework that takes two domain-specific ontologies as input, finds semantic correspondences between the concepts of both ontologies and produces a single merged ontology as output. We assume that all parties use a single language for representing ontologies. In our framework we will use OWL which is recommended by the W3C as the standard ontology language for the Web. The output of the merging process will be exploited as input for discovering and enriching the knowledge base with missing concepts and instances. Figure 1 summarizes the main steps of the proposed framework.

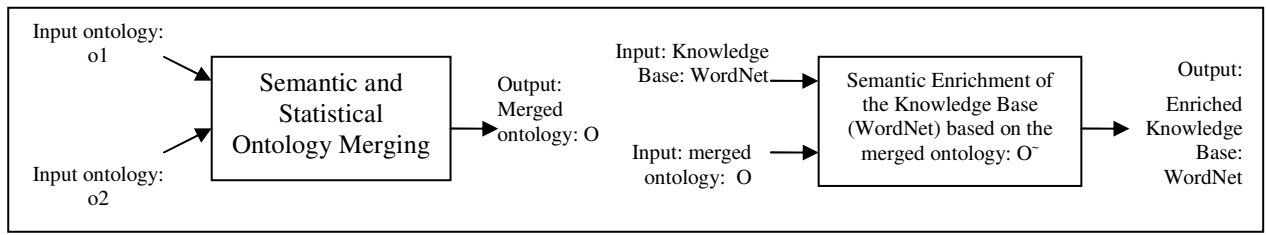

**Figure 1: Proposed Ontology Merging and WordNet Enrichment Framework**

In order to illustrate the proposed framework, we take an example of two domain-specific ontologies, Biblio.owl and BibTex.owl which were developed in the Maponto[2] project and used to describe the Bibliography domain. Figures 2a and 2b show each of the two ontologies generated by the OWL VIS tool in the Protégé[3] ontology generation environment developed at the Stanford Centre for Biomedical Informatics Research. Classes in each ontology are connected through the "is-a" partial order.

In our approach, we will use WordNet as a reference knowledge base to serve in the process of finding semantic correspondences between the concepts of the source ontologies. We choose to use WordNet because of two main reasons. First, it is a generic knowledge base and covers several domains. Second, it organizes the concepts into synsets, where each synset contains synonym concepts such as for example {writer and author}. These synsets are connected through semantic relations such as hypernymy, hyponymy, meronymy etc. This organization of WordNet synsets is useful in discovering semantic relations between the concepts of the source ontologies as we can translate the labels which represent the concepts of the ontologies into logical formulas and map them to their corresponding senses in WordNet.

On the other hand, when we have a concept or instance in any of the source ontologies which is missing in WordNet, we will submit it to another external resource in order to discover its semantics. To do so, we start from finding statistical information about the missing concept. We need to know statistically how often this concept is considered as a sub or super concept of the concepts of the source ontologies. This statistical information can be obtained from the Web, allowing us to discover the level of consensus of the web community for a specific hypernymy/hyponymy relation between two concepts. For example, for a concept like "Corporate Body" which is missing in WordNet, we need to acquire the hypernymy/hyponymy relation between it and the other concepts of the source ontologies. Based on the computed statistical results, we will infer the semantic relations between the concepts of the source ontologies and that missing concept. This step has two major benefits. First, we will be able to enrich the WordNet knowledge base with the missing concepts and instances and thus, enable the reuse of the newly discovered background knowledge in future ontology merging. Second, as this step will be fully automated, it will save time required to define the relations between the missing concepts and other concepts of WordNet.

---

[2] http://www.cs.toronto.edu/semanticweb/maponto/

[3] http://protege.stanford.edu/

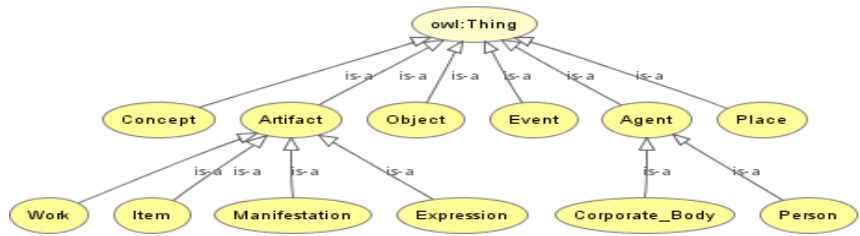

Figure 2a: Biblio Ontology

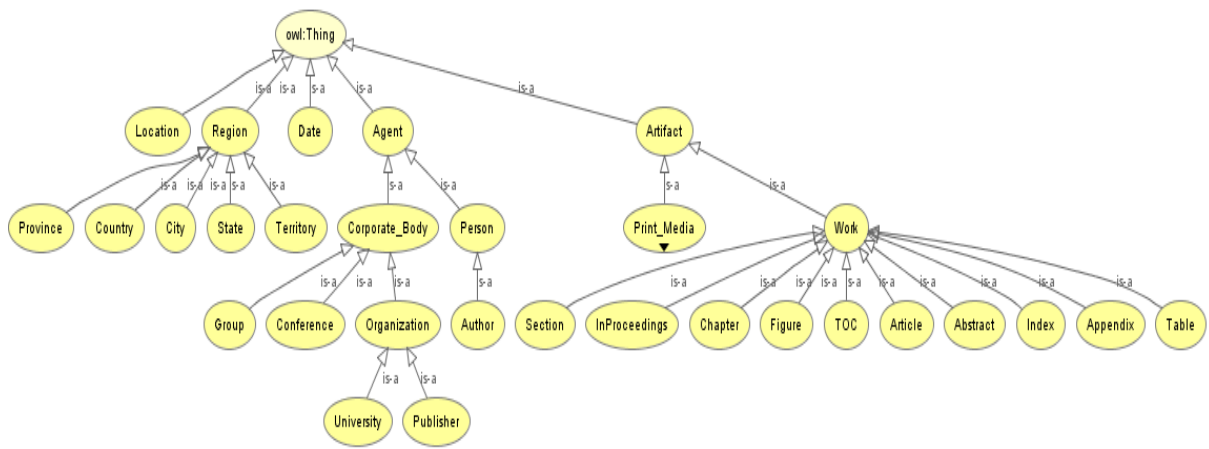

Figure 2b: Part of BibTex Ontology

## 2. RELATED WORKS

Different systems and tools have been proposed to solve the ontology semantic heterogeneity problem. These systems exploit several ontology matching strategies varying from syntactic-based to semantic-based to a combination of these approaches. In this section we give an overview of some of the existing ontology merging and alignment approaches.

**PROMPT** [15] is an algorithm for ontology merging and alignment embedded in the Protégé ontology development environment. It performs some tasks automatically and guides the user in performing other tasks for which his intervention is required. It also determines possible inconsistencies in the state of the ontology, which result from the user's actions, and suggests ways to remedy these inconsistencies. The algorithm takes two ontologies as input and guides the user in the creation of a single merged ontology as output. It also determines conflicts introduced in the ontology and finds possible solutions for those conflicts.

**FCA-MERGE** [16] was designed for merging ontologies following a bottom-up approach which offers a structural description of the merging process. The algorithm applies techniques from natural language processing and formal concept analysis in order to derive a lattice of concepts. The process of ontology merging takes as input two or more source ontologies and a set of natural language documents. The documents have to be relevant to both ontologies so that they can be described by the concepts contained in the ontologies. From the set of documents, instances are extracted and formal contexts are computed using the FCA (Formal Concept Analysis) technique which allows for a mathematical description of the concepts. For each ontology, a formal context indicates which ontology concepts appear in which documents. Then the FCA-MERGE core algorithm merges the contexts and computes a concept lattice. The final step of deriving the merged ontology from the concept lattice requires human interaction. Based on the concept lattice and the sets of relation names, the ontology engineer creates the concepts and relations of the target ontology.

The algorithm **X-SOM** [6] is used to map and integrate ontologies through homogeneous mapping where classes are mapped to classes, roles to roles, and individuals to individuals. The mapping relationships considered are equivalence and subsumption expressed through the OWL/RDFS primitives owl:equivalentClass and rdfs:subclassOf respectively. The X-SOM architecture is composed by three subsystems respectively tackling matching, mapping and inconsistency resolutions. The Matching Subsystem is constituted by a set of modules, each of which implements a specific matching technique. Each module receives as input two ontologies and returns a set of proposed mappings between pairs of resources with a similarity degree, called similarity map. Similarity maps are collected by the Mapping Subsystem and weighed by means of a neural network, in order to compute an aggregate matching value v ε [0,1] for each pair of resources. The set of candidate mappings computed by the Mapping Subsystem is handed to the Inconsistency Resolution Subsystem which is responsible for guaranteeing the global consistency of the final model.

In [1] the Ontology Matching Problem (OMP) is viewed as an n:m matching problem. Finding mappings is limited to entities of types, classes and relationships only. The degree of similarity between the entities of the input ontologies is denoted by the similarity matrix L which includes values in the range [0,1]. A matching matrix M with dimensions n x m and with entries $M[i,j] \in [0,1]$ is used to determine whether two entities are matchable or unmatchable. The proposed neighbor search algorithm has three phases. First, in the initialization phase, a partial set of similarity measures is applied to the input ontologies to determine a single initial state $St_{ini}$ for the search algorithm. In the second phase, search is in the neighborhood of the initial state. The neighbors of state $St_{ini}$ are the mapping states that can be computed either by adding to or removing from $St_{ini}$ a couple of vertices, the total number of the neighbor states will be $n*m$. In the third phase (evaluation phase), the algorithm will apply the next level(s) similarity techniques in order to find $St_f$, the best possible matching state solution.

**S-Match** [7] is an algorithm that takes two graph-like structures and produces a mapping between those nodes of the two graphs that correspond semantically to each others. The algorithm computes the semantic correspondence and returns as a result, the semantic relations implicitly or explicitly codified in the labels of nodes and arcs. Possible semantic relations are: equivalence, generalization, specialization, mismatch and overlap. The algorithm exploits several pre-processing techniques for translating the labels written in a natural language into a more precisely defined internal language. This translation is performed in order to avoid the problems related to understanding the natural language. Examples of the pre-processing techniques are tokenization, lemmatization, building atomic concepts and complex concepts. The algorithm uses string based techniques to compute the semantic relations holding between the labels at the nodes of the graphs such as postfix, prefix and n-grams. On the other hand, another resource is used to compute the semantic relations between the concepts of labels such as WordNet. This technique is a stronger semantic matcher than the string-based technique and therefore the algorithm uses it first. If WordNet fails to return the relation then other string-based techniques are used.

**IF-Map** [8] is a fully automatic ontology mapping method based on the theory of information flow proposed by [2]. The approach uses a reference ontology with no instances in order to produce an alignment structure between both local ontologies where each of them has its instances. The algorithm starts by acquiring ontologies through applying a variety of methods such as downloading existing ontologies from ontology libraries or extracting them from the Web. Then, it translates the formats of the available ontologies into Prolog clauses using different translation techniques such as RDF-to-Prolog and Ontolingua-to-Prolog. The algorithm automatically generates a logical infomorphism, that is possible mappings between both local ontologies and the reference ontology based on the information flow theory.

## 3. GENERAL OVERVIEW

To address the semantic heterogeneity problem between domain-specific ontologies, and in order to build a more complete and generic ontology, we propose a framework for *i)* Discovering and correcting the semantic relations between the concepts of the ontology in a specific domain *ii)* semantic and statistical merging of domain-specific ontologies and *iii)* WordNet enrichment based on the resulting merged ontology.

In figure 3, we consider a subset of the concepts of each of the ontologies shown in figure 2a and 2b and apply on them the steps of the proposed framework.

Before we start the merging process, we perform a check on each of the input ontologies. This check validates whether the semantic relations between the concepts of the input ontologies are conflicting with the semantic relations of the concepts in WordNet. When we find a conflict, we resolve it through reconstructing the hierarchy of the input ontology based on WordNet. This step is shown in figure 3, Step No. 1.

For example, if we take the part of Biblio ontology shown in figure 4, we find a conflict in the semantic relation between the concepts "Agent" and "Person". In WordNet, "Agent" is a hyponym of Person, while in Biblio, "Agent" is a Hypernym of "Person". Therefore, we reconstruct the Biblio ontology after computing the semantic relations between its concepts according to WordNet. The same process will be done for the second ontology, the BibTex ontology, which is shown in figure 2b.

Once we resolve the semantic conflicts for each ontology, we merge both ontologies through finding semantic relations (such as equivalence ($\equiv$), specification ($\subset$), generalization ($\supset$), disjointness ($\perp$)) between their concepts based on WordNet, and build a new merged ontology as shown in figure 3, Step No. 2. For example, the identified relations between the concepts of both ontologies are shown in table 1.

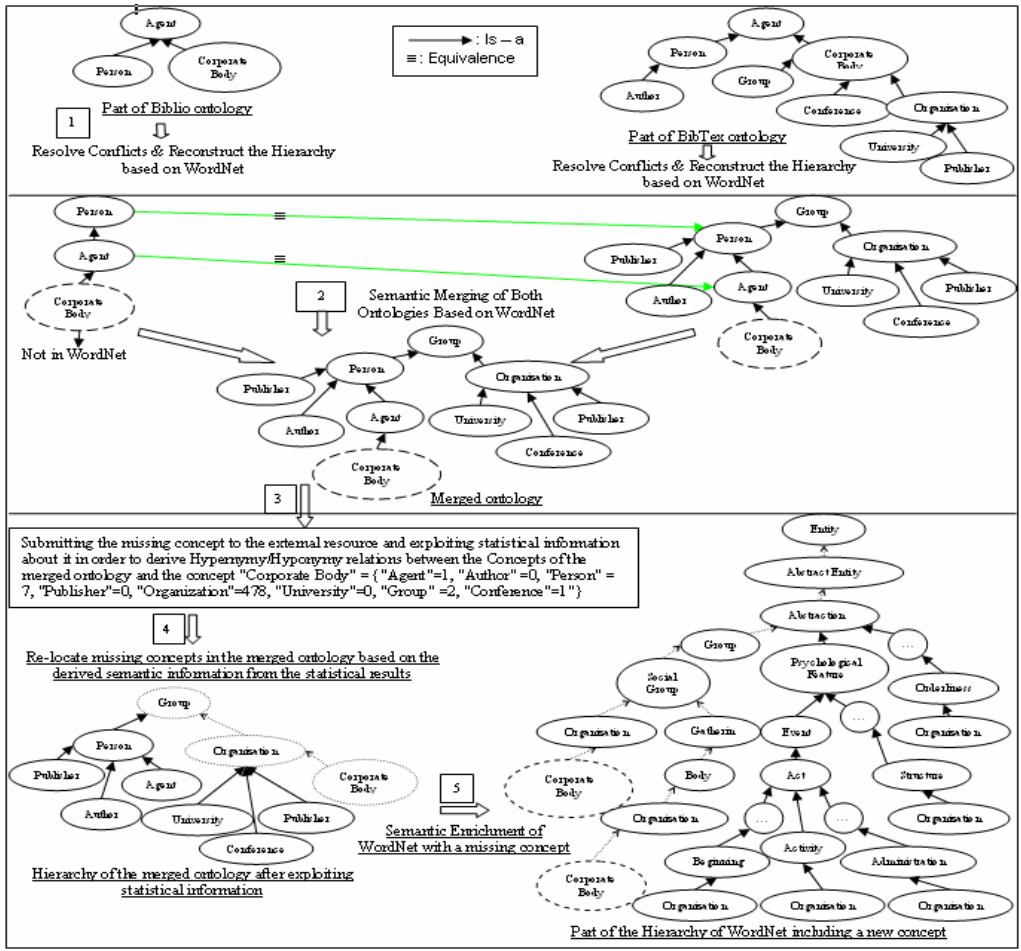

**Figure 3: Detailed steps of the proposed semantic and statistical ontology merging framework**

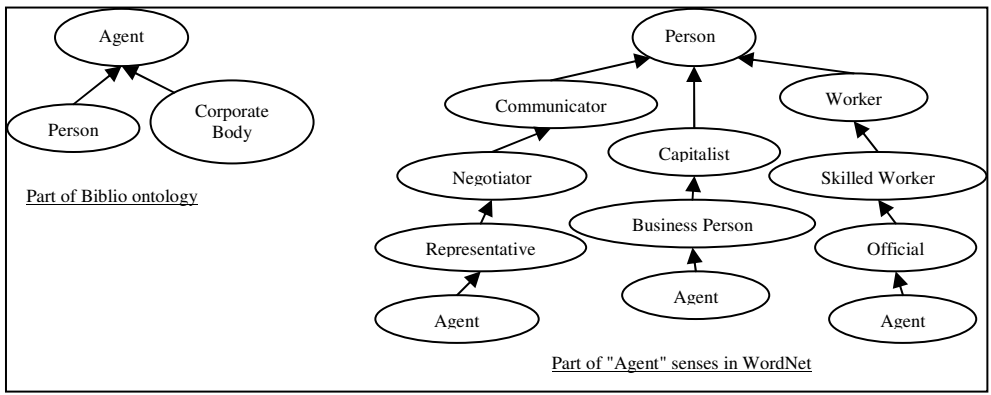

**Figure 4: conflicting semantic relations between the source ontology and WordNet**

During this step, a problem may arise when trying to compute the semantic relation between two concepts while one or both of them are not in WordNet. In this case, and as a preliminary step, we retain the relation between the missing concept and its super-concept(s) as it was in the input ontologies.

Then, we submit the missing concept to another external resource in order to compute the hypernymy/hyponymy relations between it and every concept in the merged ontology. We start by finding statistical information about the missing concept. This statistical information will be obtained from the Web and exploited in order to give an indication for the level of consensus of the web community for a specific hypernymy/hyponymy relation between the missing concept and the concepts of the merged ontology. For example, for a concept like "Corporate Body" which is missing in WordNet, we need to find the hypernymy/hyponymy relation between it and the other concepts of the merged ontology. This step is shown in figure 3 Step No. 3.

**Table 1: identified semantic relations between the concepts of both ontologies**

|               | Agent | Person | Corporate Body | Concept | Place | Object | Artifact | Expression | Item | Work | Manifestation | Event |
|---------------|-------|--------|----------------|---------|-------|--------|----------|------------|------|------|---------------|-------|
| Agent         | ≡     | ⊂      | ??             | ⊥       | ⊥     | ⊂      | ≡        | ⊥          | ⊥    | ⊥    | ⊥             | ⊥     |
| Person        | ⊃     | ≡      | ??             | ⊥       | ⊥     | ⊂      | ⊥        | ⊥          | ⊥    | ⊥    | ⊥             | ⊥     |
| Corporate Body| ??    | ??     | ≡              | ??      | ??    | ??     | ??       | ??         | ??   | ??   | ??            | ??    |
| Concept       | ⊥     | ⊥      | ??             | ⊥       | ⊥     | ⊥      | ⊥        | ⊥          | ⊥    | ⊥    | ⊥             | ⊥     |
| Place         | ⊥     | ⊥      | ??             | ⊥       | ≡     | ⊂      | ⊥        | ⊥          | ⊂    | ⊂    | ⊥             | ⊂     |
| Object        | ⊃     | ⊃      | ??             | ⊥       | ⊥     | ≡      | ⊥        | ⊥          | ⊃    | ⊥    | ⊥             | ⊥     |
| Artifact      | ⊥     | ⊥      | ??             | ⊥       | ⊥     | ⊂      | ≡        | ⊥          | ⊥    | ⊃    | ⊥             | ⊥     |
| Expression    | ⊥     | ⊥      | ??             | ⊥       | ⊥     | ⊥      | ⊥        | ≡          | ⊥    | ⊥    | ≡             | ⊂     |
| Item          | ⊥     | ⊥      | ??             | ⊥       | ⊥     | ⊂      | ⊥        | ⊥          | ≡    | ⊥    | ⊥             | ⊥     |
| Work          | ⊥     | ⊥      | ??             | ⊥       | ⊥     | ⊂      | ⊂        | ⊥          | ⊥    | ≡    | ⊥             | ⊂     |
| Manifestation | ⊥     | ⊥      | ??             | ⊥       | ⊥     | ⊥      | ⊥        | ≡          | ⊥    | ⊥    | ≡             | ⊂     |
| Event         | ⊥     | ⊥      | ??             | ⊥       | ⊃     | ⊥      | ⊥        | ⊃          | ⊥    | ⊃    | ⊃             | ≡     |

Based on the computed statistical results, we derive the semantic relations between the concepts of the merged ontology and the missing concept. And thus, we will re-locate the missing concept in the hierarchy of the merged ontology. Therefore, the concept "Corporate Body" will be considered as a sub-concept of the concept "Organization" according to the results of the statistical processing step as shown in figure 3 Step No. 4.

The final step of our approach will be to enrich WordNet with the missing concepts. This requires finding the appropriate semantic path(s) for locating it, i.e. concepts under and above which it shall be attached. To do so, we check the semantic path(s) that originate(s) from the missing concept in the merged ontology and map it/them to the proper semantic path(s) in WordNet. The result of applying this step is shown in figure 3. Step No. 5. As we can see the concept "Organization" has seven senses, but we considered only two semantic paths originating from this concept and attached the concept "Corporate Body" to these two semantic paths only.

## 4. DETAILED STEPS OF THE PROPOSED FRAMEWORK

Before we detail the steps of the proposed framework, we formalize the use of the terms "Ontology" and "Ontology Merging".

**Definition1**: **Ontology**: An ontology $O$ is *4-tuple* $\langle C, R, I, A \rangle$ where:

- $C$ represents the set of concepts of the ontology
- $R$ represents the set of semantic relations holding between the ontology concepts. They are further detailed in section 4.1.
- $I$ is the set of instances or individuals
- $A$ is the set of axioms which are first-order logic formulas.

**Definition2**: **Ontology Merging**: Given two domain-specific ontologies *o1* and *o2*, the merging operation finds semantic correspondences between their concepts and produces a single merged ontology *O* as output. Semantic correspondences between both ontologies are *4-tuples* $\langle c_{id}, a_i, b_j, r \rangle$ such that:

- $c_{id}$ is a unique identifier of the correspondence.
- $a_i \in o1$, $b_j \in o2$ are corresponding concepts of the input ontologies.
- $r$ is a semantic relation holding between both elements $a_i, b_j$.

We detail the steps of our proposed approach as follows. The first step represents a preliminary phase before the ontology merging process takes place. The second step represents the process of merging the ontologies. In the third step, when we find concept(s) missing in WordNet, we exploit statistical information about it/them in order to derive its/their semantic relation(s) with the concepts of the source ontologies. The fourth step represents the process of re-locating the missing concept in the merged ontology. And the final step is to enrich the WordNet knowledge base with the missing concept(s). We discuss below each of the proposed steps.

### 4.1 Conflict Resolution and Inconsistency Checking

At this step, for each input ontology, we find if any contradiction exists as far as semantic relations between its concepts and those in WordNet are concerned. To accomplish this task, we use the latter for computing the semantic relations between the source ontology concepts. These semantic relations are:

**Equivalence** (≡): Either concepts are equal or one of them is a synonym of the other. (E.g. Student is equivalent to Pupil).

**Specialization** (⊂): If a concept or one of its senses is a hyponym or meronym to another concept or its senses. (E.g. Student is less general than Person).

**Generalization** (⊃): If a concept or one of its senses is a hypernymy or holonym to another concept or its senses. (E.g. Transport is more general than Car).

**Disjointness** (⊥): If both concepts or their senses are antonyms or different hyponyms of the same synset. (E.g. Book is disjoint from Phone)

**Unknown** (??): If one or both of the compared concepts is/are missing in WordNet.

In case we find a semantic relation in each of the domain-specific ontologies which is contradicting with the semantic relation in WordNet, we resolve this contradiction by replacing the contradicting semantic relation with the semantic relation from WordNet.

In order to illustrate this step we take the following example. We have the concepts "Agent" and "Person" in the Biblio ontology shown in figure 2a. "Agent" is considered as a super-concept of "Person". As shown in the OWL syntax below:

<owl:Class rdf:ID="Person">

  <rdfs:subClassOf rdf:resource="#Agent"/>

 </owl:Class>

However; "Agent" is considered as a sub-concept of "Person" in the WordNet knowledge base. This case shows a contradiction between the semantic relations of both Biblio and WordNet. Therefore, to resolve this contradiction we consider the semantic relations defined in WordNet to have higher priority than those in the input ontologies. So, in Biblio, we substitute the relation between "Agent" and "Person" with the one in WordNet, i.e. we consider "Agent" as a sub-concept of "Person". The syntax will then be modified as follows:

<owl:Class rdf:ID="Agent">

  <rdfs:subClassOf rdf:resource="#Person"/>

 </owl:Class>

At this phase, we apply stop word removal and stemming techniques as a pre-processing step before submitting the concepts to WordNet. This step is a primary step of ontology merging.

## 4.2 Ontology Merging Based On WordNet

For those concepts existing in both ontologies and in WordNet, we find the semantic relations that exist between them. The process of building the hierarchy of the merged ontology will be based on the computed semantic relations between the concepts of the source ontologies. At this step, the concepts that do not exist in WordNet will be located in the merged ontology according to their locations in the source ontologies.

For example, when applying step No.2 in figure 3, we identify the following relations between the concepts of the input ontologies:

1. Agent ≡ Agent , Agent ⊂ Group, Agent ⊂ Person
2. Corporate Body ≡ Corporate Body
3. Person ⊃ Agent, Person ⊂ Group, Person ⊃ Publisher, Person ≡ Person, Person ⊃ Author
4. Group ⊃ Agent, Group ≡ Group , Group ⊃ Conference, Group ⊃ Organization, Group ⊃ University, Group ⊃ Publisher, Group ⊃ Person
5. Conference ⊂ Group, Conference ≡ Conference, Conference ⊂ Organization
6. Organization ⊂ Group, Organization ⊃ Conference, Organization ≡ Organization
7. University ⊂ Group, University ⊂ Organization, University ≡ University
8. Publisher ⊂ Group, Publisher ⊂ Organization, Publisher ⊂ Person, Publisher ≡ Publisher
9. Author ⊂ Person, Author ≡ Author

Then, the hierarchy of the merged ontology is constructed based on the computed semantic relations between the concepts of both ontologies.

## 4.3 Exploiting Statistical Information about Missing Concepts

Our source of acquiring the relation(s) between the missing concept(s) and other concepts of the merged ontology is the massive amount of information encoded in texts on the Web. This textual information represents facts, people's opinions, and ideas. We know that information encoded in texts on the web is subjective and sometimes may represent incorrect semantic information. Therefore, in our approach we limit subjectivity and reduce the chance of extracting incorrect semantic information by only acquiring relations between the concepts of the merged domain-ontologies and those which are missing in WordNet.

For example, to extract a hyponymy relation between two concepts, we build on the definition of this relation in [14]: a concept represented by the synset {x, x', …} is said to be a hyponym of the concept represented by the synset {y, y', …} if native speakers of English accept sentences constructed from such frames as "An x is a (kind of) y".

Therefore, in order to extract the hyponymy/hypernymy relation between each missing concept and the concepts of the merged ontology, we will submit the following query to the external resource "Google Search" in our case:

$Q_i$ = "($C_{missing}$) is a(n) ($C_i$)" where,

· $C_{missing}$ is the missing concept in WordNet.

· $C_i$ is a concept which exists in the merged ontology and in WordNet.

For each missing concept, queries $Q_i$ will be submitted including every concept $C_i$ in the merged ontology. Let us note that queries that include the negation operator such as "No $C_{missing}$ is a(n) $C_i$" are excluded.

For example, to acquire the hypernyms of the concept "Corporate Body" we submit the following Web pattern-based queries:

$Q_1$= *"Corporate Body is an Organization"* which outputs 478 results (res_$Q_1$=478).

$Q_2$= *"Corporate Body is a Publisher"* which outputs 0 results (res_$Q_2$=0).

Based on the obtained results of queries $Q_i$, the concepts $C_i$ that co-occur more frequently with the missing concept are then suggested as hypernyms. This step is based on using an automatic threshold value $t$ to automatically decide which concepts among the list of retrieved concepts are suggested as hypernyms of the missing concept.

The value $t$ is determined based on the following process:

1. For each missing concept, after submitting queries $Q_i$ we have the following list of results:

$$res\_total = [res\_Q_1, res\_Q_2, res\_Q_3, ..., res\_Q_n]$$

2. We find the maximum difference between the number of retrieval results as follows:

$$t = maxDiff(res\_Q_i, res\_Q_j)$$

3. Based on $t$, a sub-list of the list of concepts is then suggested comprised of the candidate hypernyms of the missing concept. If res_$Q_i$ is greater than $t$ then the concept $C_i$ is suggested.

Based on the provided suggestions from the previous step, we will re-locate the missing concepts in the merged ontology and update its hierarchy. For example, it was suggested to consider the concept "Corporate Body" as a sub-concept of the concept "Organization".

### 4.4 WordNet Enrichment

We will attach the missing concepts to the WordNet hierarchy according to the semantic path(s) that originate from them in the merged ontology. Enriching the WordNet knowledge base with a new concept requires finding the appropriate position for locating it. To do so, we will check the path(s) that originate(s) from the missing concept in the merged ontology and map it/them to the proper path(s) in WordNet. Below we summarize the different cases that we consider for the enrichment.

**Case1:** We have the concept C in the merged ontology, and C is a sub-concept of one direct super-concept X. The concept X has only one sense in WordNet, i.e. there is only one semantic path that originates from X.

In this case, we locate the missing concept C under its direct super-concept X in the semantic path that originates from X in Wordnet. For example, the term "Concept" in WordNet has only one sense. Therefore, any term considered as a sub-concept of this concept will be directly located under it. Figure 5a below shows the result of applying **Case1**.

**Case2:** We have the concept C in the merged ontology, and C is a sub-concept of one direct super-concept X. The concept X has more than one sense in WordNet, i.e. there is more than one semantic path originating from X.

In this case, based on the merged ontology and WordNet, we check the semantic similarity between the list of concepts in the semantic path originating from the concept X in the merged ontology and the list of concepts in the semantic paths originating from the senses of X in WordNet. Then, for the most similar semantic path(s) that originate from the concept X, we locate the missing concept C as a sub-concept of the concept X.

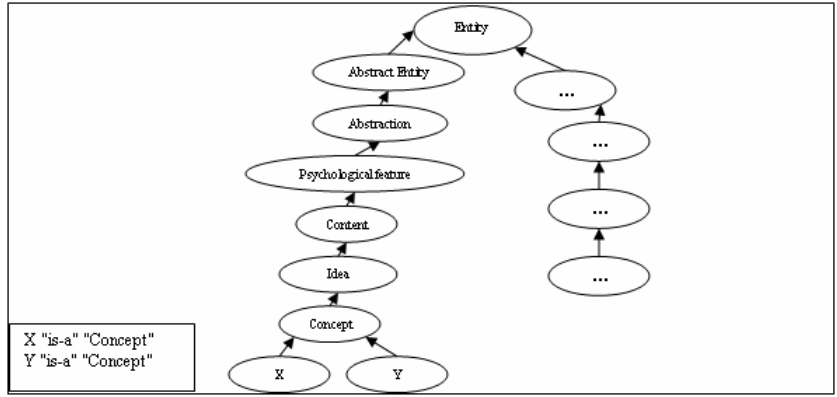

**Figure 5a: Result of applying Case1**

Figure 5b illustrates the case of enriching WordNet with the concept "Corporate Body". As we can see in the merged ontology, this concept is a sub-concept of the concept "Organization". In WordNet, the concept "Organization" has more than one sense. Therefore, we need to find which sense among the several senses of this concept will be considered as a super-concept of "Corporate Body". This step can be done based on the comparison between the semantic path which originates from "Organization" in the merged ontology and those originating from "Organization" in WordNet. The semantic path(s) most similar to the one in the merged ontology will be considered in order to attach the concept "Corporate body" to them.

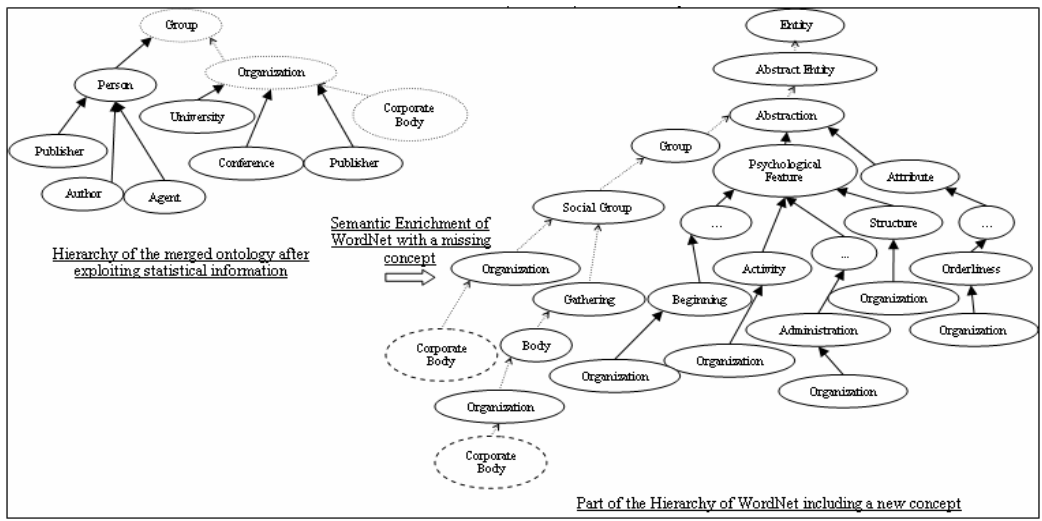

**Figure 5b: Result of applying Case2**

**Case3:** We have the concept C in the merged ontology which is a sub-concept of more than one direct super-concept. In this case, we check whether either Case1 or Case2 applies for each direct super-concept of the missing concept C.

## 5. EXPERIMENTAL RESULTS

### 5.1 Experiments Using our *Expert Mappings*

We used the Biblio and BibTex ontologies, both covering the bibliographical domain, in our experiments. They have a different number of concepts: there are twelve concepts in the Biblio Ontology and forty-three concepts in BibTex. In order to provide a ground for evaluating the proposed techniques, we manually identified all possible alignments between the concepts of both ontologies. We identified alignments between the concepts of both ontologies based on our knowledge and experience in order to produce *expert mappings* in the same fashion as the mappings highlighted in [7] and [18]. Then, we compare these expert mappings with the ones produced by the system automatically. The experiments we perform are classified into the following categories:

1. Merge the ontology to itself, i.e. self-merging
2. Merge both ontologies together

### 1. Self-merging

## *1.1 Manual Merging*

In this test, the Biblio ontology is merged to itself. Alignments are identified between the concepts of this ontology. The semantic relations considered between the ontology concepts are: Equivalence (≡), Disjointness (⊥), Hypernymy or Holonymy (⊃) and Hyponymy or Meronymy (⊂).

We detail below the steps of this test:

1. We first find alignments in the ontology. In this step, it is clear that for all concepts, each one is equivalent to itself. On the other hand, when the concepts are not similar, we identify other types of semantic relations between them, such as hypernymy, hyponymy and disjointness.

2. Based on the results of the identified semantic relations between the concepts of the ontology, we build the merged ontology. The result of this step is shown in figure 6a:

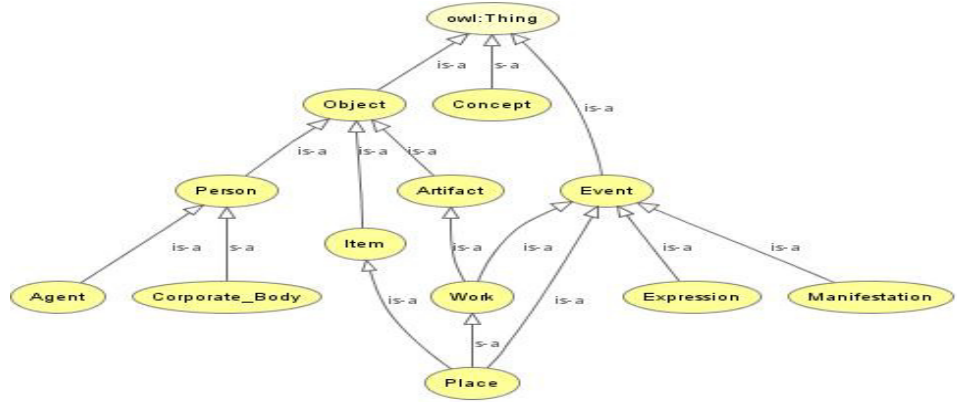

**Figure 6a: Result of manual merging of the ontology to itself**

## *1.2. Automatic Merging*

Alignments are automatically identified using our proposed ontology merging framework. At this phase, the process of identifying alignments and semantic correspondences between the concepts of the ontology consists of the following main steps:

1. First, the ontology OWL file is parsed and concepts of the ontology extracted.
2. Based on WordNet, the algorithm computes semantic relations between the concepts of the ontology.
3. It is possible to have concepts in the ontology which are not in WordNet. For example, "Corporate Body" is not in WordNet. In this case, for those concepts, the algorithm applies an automated knowledge acquisition step in order to find the semantic relations between these missing concepts and other concepts of the ontology. The statistical knowledge acquisition step is done according to the process detailed in Sec. 4.3.
4. The hierarchy of the merged ontology is built based on the results of the discovered semantic relations between the concepts of the ontology. These results are produced when applying steps 2 & 3.
5. The algorithm generates an OWL file representing the merged ontology. Figure 6b shows the resulting merged ontology.

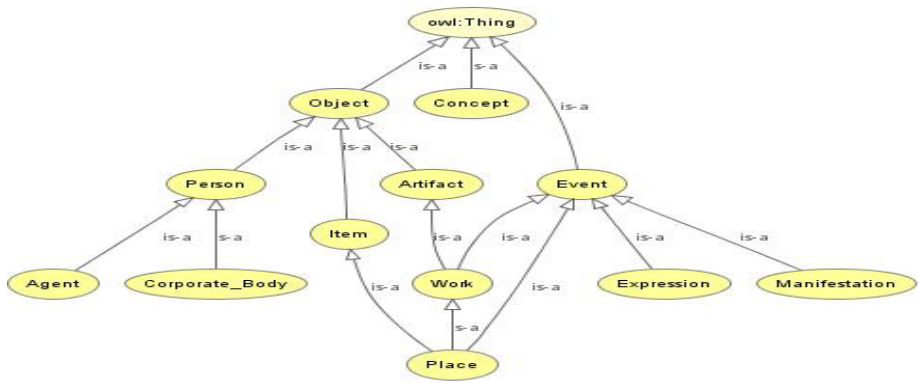

**Figure 6b: Result of automatically merging the ontology to itself using our proposed framework**

## 2. Merging two ontologies from the same domain

### 2.1 Manual Merging
At this step, we identify alignments through measuring the semantic relations between the concepts of both the Biblio and BibTex ontologies. These semantic relations are the same as those described in the first test.

We manually build the hierarchy of the merged ontology based on the defined semantic relations between the concepts of both ontologies. Figure 7a shows the hierarchy of the merged ontology.

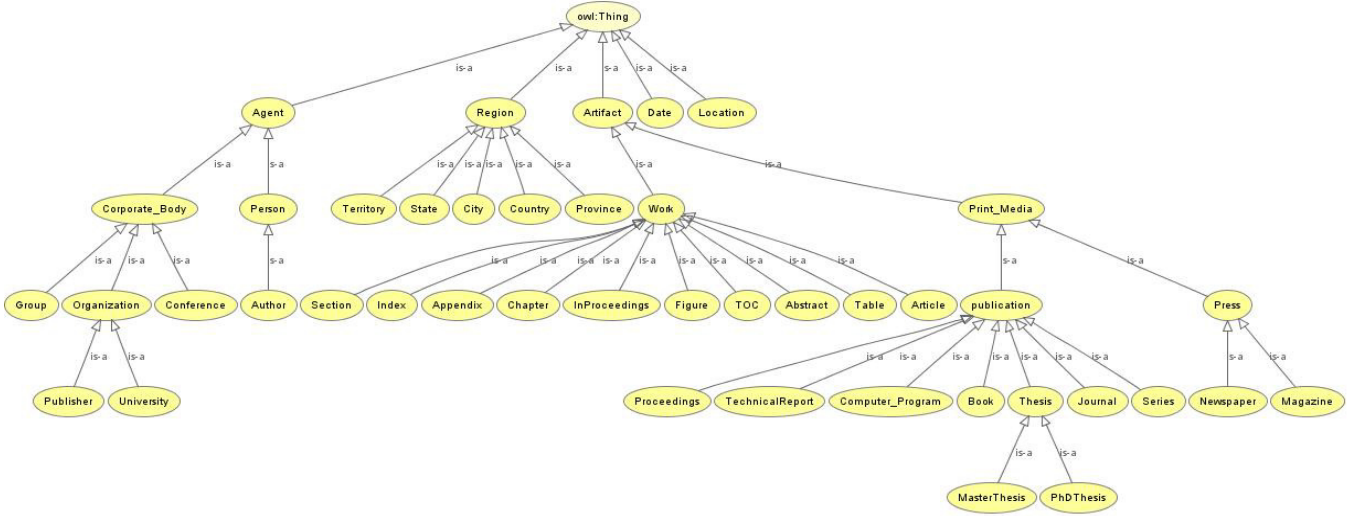

**Figure 7a: Result of manually merging two domain-specific ontologies**

### 2.2 Automatic Merging
We apply the same steps as those described in Sec. 1. in order to merge both the Biblio and BibTex ontologies.

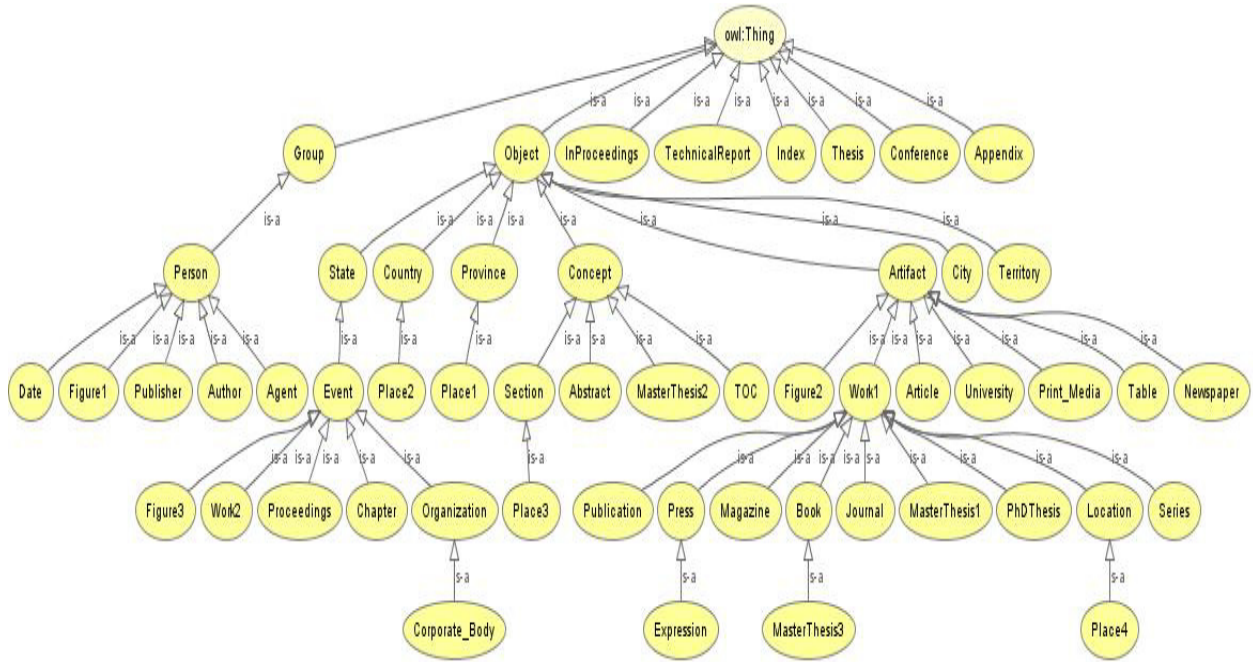

**Figure 7b: Result of automatically merging two domain-specific ontologies**

At step 3, the algorithm identifies the followings concepts which do not exist in WordNet:
1. Corporate_Body
2. InProceedings

3. TechnicalReport
4. TOC
5. MasterThesis
6. PhDThesis

Semantic relations between these concepts and other concepts of both ontologies can therefore not be computed. Consequently, at this step, the algorithm applies the automated knowledge acquisition module routine in order to acquire the semantic relations between these concepts and other concepts of both ontologies. Figure 7b shows the resulting merged ontology.

We use the precision and recall indicators in order to measure the quality of our results. Table 2 shows the results of this step:

**Table 2: Precision/ Recall results for our expert mappings**

| Test | P | R |
|---|---|---|
| Test1. Self-Merging | 1.0 | 1.0 |
| Test2. Merge two domain-specific ontologies | 1.0 | 0.94 |

## 6. EXPERIMENTS USING EXPERT MAPPINGS USED BY [7] & [18]

We use in these experiments the expert mappings[4] which were manually produced as detailed in [7]. Manual matches identified between the attributes of the ontologies were not considered in these tests.

In these tests, we combine both semantic and statistical automatic knowledge acquisition techniques in order to find alignments and merge both ontologies. The semantic technique is given the priority and if it fails we use the statistical automatic knowledge acquisition technique to acquire hypernymy-hyponymy relations between the concepts of both ontologies and the concepts that are missing in WordNet.

We use four pairings of ontologies: Parts of Google and Yahoo web directories, Product Schemas, Simple Catalogs, and Company profiles. The mappings produced by our techniques are classified into: i)correct, where the produced semantic relations between the concepts of both ontologies are correct and exist in the expert mappings; ii)incorrect, where the prosecuted semantic relations are not correct; and iii)others:, where the produced semantic relations between the concepts of both ontologies are correct but do not exist in the expert mappings.

We compared the results produced by our techniques with the results from the expert mappings as shown in table 3.

**Table 3: Expert mappings and system results**

| Ontologies | Expert mappings | Correct | Not correct | Others |
|---|---|---|---|---|
| Google and Yahoo Web Directories (9 terms * 6 terms = 54Comparison) | 4 | 4 | 0 | 6 |
| Product Schemas (5 terms * 6 terms = 30Comparison) | 4 | 4 | 0 | 0 |
| Company profiles (3 terms * 3 terms = 9 Comparison) | 3 | 2 | 0 | 0 |
| Simple Catalogs (2 terms * 3 terms = 6 Comparison) | 3 | 3 | 0 | 0 |

Then, we compare our results with three state-of-the-art syntactic-based and/or semantic-based systems: Cupid [12], COMA [3], and S-Match [7]. In table 4, we show our comparative results with the above mentioned systems. In these results, we consider the attributes of both ontologies in order to measure precision and recall. The number of terms in our test ontology is 20 terms (4 terms in the first ontology and 5 terms in the second ontology).

---
[4] http://dit.unitn.it/~accord/Experimentaldesing.html

As shown in Table 4, our framework shows higher precision and recall than both the COMA and Cupid systems. It is furthermore comparable to the S-Match system.

Table 4: Precision/ Recall results for our system and other matching systems

| Ontology | Cupid | | COMA | | S-Match | | Proposed Framework | |
|---|---|---|---|---|---|---|---|---|
| | P | R | P | R | P | R | P | R |
| Simple Catalogs (20) | 0.44 | 0.36 | 0.62 | 0.66 | 1.0 | 1.0 | 1.0 | 1.0 |

## 7. CONCLUSION AND FUTURE WORKS

We have introduced a fully-automated framework for merging domain-specific ontologies and WordNet enrichment through exploiting both semantic and statistical-based techniques. On the one hand, this paper applies a semantic-based technique to merge domain-specific ontologies through finding semantic correspondences among their concepts. And on the other hand, when the semantic relation between two concepts cannot be identified (because one of them or both are not in WordNet), it applies a statistical-based technique to *i)* derive the semantic relation between the undefined concepts and other concepts in the merged ontology and *ii)* to enrich WordNet (which we use to support the process of ontology merging) with the missing concepts.

Future work will include applying our method on other types of relations among the concepts of the ontologies. We also plan to plug in other matching techniques such as structural techniques in order to enhance our framework.